\documentclass{aastex}          
\usepackage{spr-astr-addons}    
\usepackage{natbib}

\usepackage{color}

\begin{document}
\title{Paired galaxies with different activity levels and their supernovae}

\shorttitle{Paired galaxies with different activity levels and their supernovae}
\shortauthors{Nazaryan~et~al.}

\author{\small{T.A.~Nazaryan}}
\email{nazaryan@bao.sci.am} 
\and
\author{\small{A.R.~Petrosian}}
\and
\author{\small{A.A.~Hakobyan}}
\affil{Byurakan Astrophysical Observatory, 0213 Byurakan,\\
Aragatsotn Province, Armenia,\\
e-mail: nazaryan@bao.sci.am
}
\and
\author{\small{V.Z.~Adibekyan}}
\affil{Centro de Astrof\'{i}sica da Universidade do Porto,\\
Rua das Estrelas, 4150-762 Porto, Portugal}
\and
\author{\small{D.~Kunth}}
\and
\author{\small{G.A.~Mamon}}
\affil{Institut d'Astrophysique de Paris, UMR 7095 CNRS-UPMC,\\
98bis Bd Arago, 75014 Paris, France}
\and
\author{\small{M.~Turatto}}
\affil{INAF-Osservatorio Astronomico di Padova,\\
Vicolo dell'Osservatorio 5, 35122 Padova, Italy}
\and
\author{\small{L.S.~Aramyan}}
\affil{Byurakan Astrophysical Observatory, 0213 Byurakan,\\
Aragatsotn province, Armenia}

\begin{abstract}
We investigate the influence of close neighbor galaxies on the properties of supernovae (SNe)
and their host galaxies using 56 SNe located in pairs of galaxies
with different levels of star formation (SF) and nuclear activity.
The statistical study of SN hosts shows that there is no significant difference between
morphologies of hosts in our sample and the larger general sample of SN hosts in the
Sloan Digital Sky Survey (SDSS) Data Release~8 (DR8).
The mean distance of type II SNe from nuclei of hosts is
greater by about a factor of 2 than that of type Ibc SNe.
The distributions and mean distances of SNe are consistent
with previous results compiled with the larger sample.
For the first time it is shown that SNe Ibc are located
in pairs with significantly  smaller difference of radial velocities
between components than pairs containing SNe Ia and II.
We consider this as a result of higher star formation rate (SFR) of these closer systems of galaxies.
SN types are not correlated with the luminosity ratio of host and neighbor galaxies in pairs.
The orientation of SNe with respect to the preferred direction toward neighbor galaxy
is found to be isotropic and independent of kinematical properties of the galaxy pair.
\end{abstract}


\section{Introduction}

Stellar populations and history of SF are crucial parameters that determine the nature of galaxies.
Among other triggers, interactions and mergings can play an important role in explaining
the origin and processes underlying the active SF phenomena observed in galaxies
\citep[e.g.,][]{blanton09,bournaud11}.
Particularly, merging is considered as the mechanism that changes the properties of galaxies
dramatically and is able not only to bring additional amount of gas
for fueling nuclear activity and starburst, but also to change
the efficiency and timescales of SF in galaxies significantly \citep[e.g.,][]{kennicutt98,cox08}.
Connection between interaction/merging and SF has been
shown in many observational studies \citep[e.g.,][]{barton00,patton11} and theoretically has
been modeled and explained in many papers \citep[e.g.,][]{mihos96,dimatteo07}.
According to large-sample statistical studies,
in most cases gravitational interaction can be a triggering mechanism
for nuclear activity and/or circumnuclear starburst
in interacting and merging galaxies \citep[e.g.,][]{ho08,ellison11,liu12, patton11,patton13}.
In several studies enhanced SF was revealed also in tidal arms and in bridges
connecting galaxies in interaction \citep[e.g.,][]{smith07}.
There also exists the possibility, that interactions can enhance SF in disk,
while the nucleus is not undergoing even a modest SF phase \citep[e.g.,][]{jarrett06}.

SNe are classified into two main types with differences in the
nature of progenitors and mechanisms of explosion.
Current understanding is that type Ib, Ic and II SNe occur from
the gravitational collapse of young massive stellar cores \citep[][]{hamuy03,smartt09},
and type Ia SNe from thermonuclear explosions of a white dwarf in close binary systems
\citep[][]{livio01,maoz12}.

\defcitealias{H09}{H09}
Binary companions in type Ia events have unclear nature \citep[][]{mannucci05,maoz12},
whereas CC SNe are connected only with massive young stellar populations with possible
differences in masses of progenitors from less massive ones ($M \gtrsim 8M _\sun$)
for type II SNe to more massive ones ($M \gtrsim 20M _\sun$) for Ibc\footnote{\footnotesize
{The notation type Ibc SNe is often used to indicate globally
the entire class of stripped-envelope SNe, i.e Ib, Ic and intermediate cases.}}
SNe, with these boundaries depending also on metallicity \citep[e.g.,][]{boissier09}.
While SNe Ia are found in any morphological types of galaxies \citep[][]{vandenbergh05},
CC SNe are located only in spiral and irregular galaxies
with the few exceptions of being found in early-type galaxies,
which are in close interaction or are remnants of merging processes \citep[][]{hakobyan08a}.
SN rates of both type Ibc and type II SNe
increase from early- to late-type spiral galaxies \citep[e.g.,][]{mannucci05,hakobyan11}
and are correlated with color and far-infrared (FIR) excess of galaxies \citep[e.g.,][]{cappellaro99}.
The radial distribution of Ibc and II SNe in galactic disks is exponential,
being more concentrated for type Ibc SNe
(\citealt[][]{vandenbergh97}; \citealt[][hereafter H09]{H09}; \citealt[][]{habergham12}).
CC SNe also are closely associated with spiral arms \citep[e.g.,][]{maza76}
and HII regions \citep[e.g.,][]{bartunov94,anderson08}.
All observational results mentioned above suggest that CC SNe are tightly connected with recent SF.

Several studies have been carried out to investigate rates and distributions of CC SNe
located in galaxies showing different levels of nuclear and starburst activity
\citep[][]{petrosian90,bressan02,petrosian05,hakobyan08b}.
Samples inclu\-ded both isolated and paired as well as groups of galaxies \citep[][]{petrosian95,navasardyan01}.
According to these studies, the radial distribution of SNe in active and star-forming galaxies
shows a higher concentration toward the center of the active hosts than in normal ones,
and this effect is more pronounced for CC SNe \citep[][]{petrosian05},
especially for SNe Ibc \citep[][]{hakobyan08b}.
Study of SNe in morphologically disturbed hosts has been carried out also
and has showed that    fraction of type Ibc SNe in central regions of disturbed hosts is increased
compared to that of the undisturbed hosts \citep[][]{habergham12}.
There is also an indication that SN rates are higher in galaxy pairs
compared with that in groups, which can be related to the enhanced SFR
in strongly interacting systems \citep[][]{navasardyan01}.
Studies of SNe in interacting hosts are difficult
because of small statistics and many selection effects,
but they can provide additional helpful information to understand the
influence of interactions on activity and SF processes of galaxies.

The aim of this study is to investigate to what extent
gravitational interaction with a close neighbor can be connected
with nuclear activity and/or enhanced SF in galaxy pairs,
using SNe as tracers of recent SF.
For that purpose we selected samples of paired galaxies
with different nuclear/starforming activity levels,
from AGNs and starburst galaxies to completely passive ones.
We examined possible correlations between kinematical properties of pairs of galaxies,
integral parameters of SN hosts and their neighbors,
as well as SN types and their distributions.
Section~\ref{sample} presents the sample of close pairs of galaxies with at least one SN event.
Section~\ref{method} discusses the parameters of SNe, their hosts and close neighbors.
Section~\ref{statistics} presents the statistical study of the sample and discusses its results.
Section~\ref{summary} is the summary of this study.
Throughout this paper, we adopt the Hubble constant $H_0=73 \,\rm km \,s^{-1} \,Mpc^{-1}$.

\section{Sample}
\label{sample}

\defcitealias{H12}{H12}
The sample of the current study was obtained by cross-matching
the catalog of SNe by \citet[hereafter H12]{H12}
with the sample of selected pairs of galaxies (see below).
SN database \citetalias{H12} contains 3876 SNe  located
in the area of the sky covered by the SDSS DR8 survey.
The database, with the revision of a few SNe \citep[][]{aramyan13},
particularly provides SN types, their exact positions in the hosts,
as well as homogeneous information about hosts of these SNe
(galaxy name, exact coordinates, redshift, morphology,
magnitude, angular size etc.) when available.

We used three catalogs of galaxies with different levels of nuclear activity to construct
our sample of close pairs of galaxies. These catalogs are the following:
(1) the catalog of Markarian (MRK) galaxies
(\citealt[][hereafter P07]{P07}),
(2) the Second Byu\-rakan Survey (SBS)\, galaxies catalog
(\citealt[][hereafter G11]{G11})\,
and (3) the North Galactic Pole \,(NGP)\, galaxy catalog
(\citealt[][hereafter P08]{P08}).
\defcitealias{P07}{P07}
\defcitealias{G11}{G11}
\defcitealias{P08}{P08}

The MRK catalog contains 1545 galaxies having starburst properties and/or active nuclei.
In \citetalias{P07}, homogeneously measured parameters of MRK galaxies,
such as magnitudes, sizes, positions, redshifts, morphologies are presented.
The SBS catalog contains 1676 galaxies, most of which are known to have active nuclei and/or starbursts.
Parameters of SBS galaxies are presented in a recent version of the catalog in \citetalias{G11}.
They are measured in the same way as those of the MRK sample.
The NGP catalog in \citet[][]{huchra90} contains a complete sample of galaxies
down to $15.5$ photographic magnitudes within a 6-degree strip passing through the NGP.
In \citetalias{P08}, 1093 galaxies of this catalogue were studied,
by being grouped into two samples according to their activity level.
We named these two samples accordingly as:
NGPA -- for galaxies with active or star forming nuclei,
and NGPN -- for normal galaxies.
Properties of NGPA and NPGN galaxies were also measured
in the same homogeneous way as for MRK and SBS samples.
Median redshifts for MRK, SBS and NGP galaxies are $0.02$, $0.04$ and $0.03$ and blue apparent magnitudes
with their $\sigma$-s are $15.2\pm0.9$, $16.7\pm0.9$ and $15.0\pm0.7$, respectively.
Therefore, the galaxies in three samples are located at similar redshifts
and have magnitudes close to each other.
It is important to note that, the MRK and SBS samples could bias the total sample toward more active galaxies,
both AGN hosts and starburst galaxies.
To check it, we conducted a binomial test (comparison of percentages of AGN and starforming hosts)
of SN hosts between subsamples from MRK+SBS and NGP pairs
and obtained that the proportions of AGN and starforming hosts are
the same for these two subsamples at $P=0.08$ and $P=0.15$ respectively.
Thus, MRK and SBS samples do not bias our further statistics.

\defcitealias{N12}{N12}
Results of a close neighbors search for MRK galaxies within position-redshift space
using NASA/IPAC Extragalactic Database (NED) and SDSS DR8
(for current study we added to this list some new objects
found only in the 9th data release \citep[][]{ahn12} of SDSS)
are already published (\citealt[][hereafter N12]{N12}).
In \citetalias{N12}, three criteria were used to select the sample of close neighbors of MRK galaxies.
(1) Redshift of MRK galaxy should be more than $0.005$.
(2) Difference of radial velocities of MRK galaxy and its neighbor should be less than ${\rm 800~km~s^{-1}}$.
(3) Projected distance between MRK galaxies and their neighbors should be less than 60~kpc (close systems).
According to these criteria, 633 galaxies in close systems containing 274 MRK galaxies were discovered in \citetalias{N12}.
We also conducted the search of neighbors for SBS and NGP galaxies using the same criteria as for MRK galaxies.
Identification revealed 380 galaxies in close systems containing
228 SBS galaxies, 365 galaxies in close systems containing 166 NGPN galaxies
and 420 galaxies in close systems containing 186 NGPA galaxies.
For the current study, only pairs of galaxies were selected from the
above mentioned 1798 galaxies in close systems with different multiplicity.
The total number of pairs containing at least one MRK, SBS, NGPA or NGPN galaxy is 675.
The percentage of pairs that are located within the SDSS area is 87~\%, 94~\% and 100~\% for MRK, SBS and NGP pairs, respectively.

The sample of 675 pairs of galaxies was cross-matched
with the list of SN hosts from \citetalias{H12}.
In total 56 SNe in 44 hosts were identified:
18 SNe of them located in 10 MRK, 4 SNe in 4 SBS, 3 SNe in 3 NGPN,
11 SNe in 10 NGPA hosts, and 20 SNe in 17 neighbors of the aforementioned galaxies.
Although the \citetalias{H12} database contains information about the hosts of SNe,
we once more visually inspected all our SNe and host pairs
and carefully checked the identification of host galaxies and SN offsets.

\section{Measured and collected parameters}
\label{method}

In order to obtain homogeneous measurements of parameters of all sample galaxies in pairs,
we measured them in the same way as in \citetalias{P07}, \citetalias{P08}, \citetalias{G11}, and \citetalias{N12}.
Table~\ref{listofSn} presents the data of SNe, their hosts, and host neighbors,
which were used in statistical research.
The first four columns of Table~\ref{listofSn} present data for SNe.
Column~1 contains SN names and Col.~2 contains SN types \citepalias{H12}.
We grouped SNe into three main classes: type Ia, type II, and type Ibc for statistics.
In a few cases, marked by ``*'', types have been inferred from the light curves.
Uncertainties in SN type are marked by ``:'' and ``?'' (highly uncertain).
Nineteen out of 56 SNe are of type Ia, 12 are of type Ibc, and 15 are of type II.
Column~3 contains radial distances of SNe from their host nuclei,
normalized to $R_{25}$ of hosts according to \citetalias{H09}.
For CC SNe, we assumed that they are located within the discs of the hosts and
corrected their offsets, taking into account galaxy inclination.
For SNe of type Ia, the radial distance is the ratio
of its angular distance from nuclei to the  $R_{25}$ of the host.
We did not deproject radial distances for SNe Ia,
because 2 out of 19 SNe Ia are located in early-type hosts,
6 SNe are visually located outside the disks of the hosts.
Even assuming that some of SNe Ia in spiral hosts belong
to the disk population \citep[e.g.,][]{mannucci05,mannucci06,hakobyan11},
doing inclination correction to deproject them seems likely to add more errors than if we do not deproject them\footnote{\footnotesize
{However, in the following we will make comparison
with radial distances of SNe Ia computed in similar way.}}.
Column~4 presents the position angle (PA) of the SN in its host
with respect to the direction toward the neighbor galaxy.
It is an angle that has a value from $0^{\circ}$ for a SN located toward
the neighbor galaxy to $180^{\circ}$ for a SN located on the opposite side of the host,
independently of the clockwise/anticlockwise direction.

The next 13 columns of Table~\ref{listofSn} present
integral parameters for SN hosts and their neighbors in pairs
used in our statistical analysis.
Redshifts of galaxies in pairs were obtained from the SDSS DR9 and/or NED.
Then, recession velocities were corrected for Virgocentric infall \citep{terry02}.
The median distance of pairs is 62~Mpc.
Coordinates of galaxies were obtained using peak surface brightness from the SDSS images.
Names of hosts and their neighbors are presented in Cols.~5 and 6.
The SDSS RGB images were used as a primary source
for SN hosts and their neighbors' morphological classification (Cols.~7-10).
A detailed description of classification procedure is available in \citetalias{P07}.
For checking consistency, we compared our SN hosts' morphological classifications with that of \citetalias{H12}.
The mean absolute difference in \emph{t}-types of these two classifications
is 0.6~units, and bar detection is different in 17~\% of cases.
To be consistent with \citetalias{P07}, \citetalias{P08}, \citetalias{G11}, \citetalias{N12},
magnitudes of the galaxies are measured
from the Second Palomar Observatory Sky Survey (POSS-II) and
UK Schmidt Telescope (UKST) photographic plates, collected in the Second Digitized Sky Survey (DSS-II).
Blue ($J_{\rm pg}$) and red ($F_{\rm pg}$) magnitudes of galaxies were measured
from the $J$ and $F$ band images of the objects in a homogeneous way
at the isophote corresponding to $\ge 3{\rm \sigma}$ background noise,
which is approximately ${\rm 25.2\,\ mag\,\ arcsec^{-2}}$ (see \citetalias{P07} for details).
Absolute magnitudes are corrected for Galactic foreground (\citet[][]{schlafly11},
recalibration of the \citet[][]{schlegel98} infrared-based dust map)
and target galaxy internal extinctions \citep[][]{bottinelli95}
and are presented in Cols.~11 and 12.
Linear major diameters of SN hosts and their neighbors
are presented in Cols.~13 and 14.
They are calculated using angular major diameters
measured in a homogeneous way from the DSS-II blue images (see \citetalias{N12} for details).

\begin{figure}[t]
\begin{center}
\includegraphics[width=0.74\hsize,angle=-90]{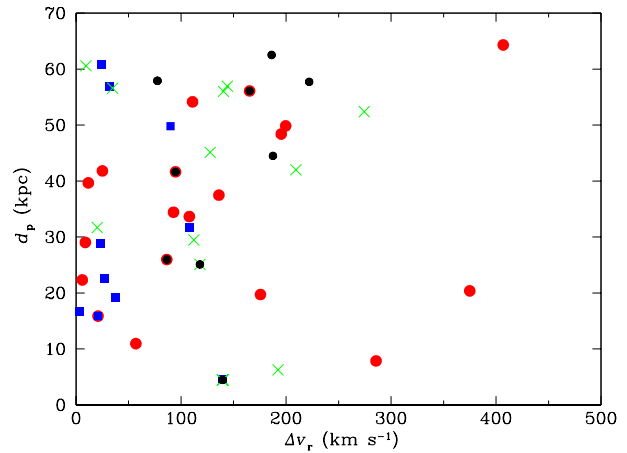}
\end{center}
\caption{The distribution of pairs according to the differences of radial velocities
and projected distances of components. The \emph{symbols} correspond to hosts with different types of SNe:
Ia (\emph{red filled circles}), Ibc (\emph{blue filled squares}), II (\emph{green crosses}),
I and unclassified SNe (\emph{black dots})}
\label{distribofPairs}
\end{figure}

Pairs of galaxies are described via two parameters describing
strength/stage of interacting/merging \citep[e.g.,][]{patton00}:
difference of radial velocities $\Delta v_{\rm r}$ of SN hosts and their neighbors (Col.~15)
and linear projected distance $d_{\rm p}$ (Col.~16) between pair members.
Figure~\ref{distribofPairs} shows the distribution of pairs according to these two parameters.
It is noteworthy, that although we cut the initial sample of pairs
by ${\Delta v_{\rm r} \leq \rm 800~km~s^{-1}}$,
all pairs hosting SNe have ${\Delta v_{\rm r} \lesssim \rm 400~km~s^{-1}}$.
The mean error for $\Delta v_{\rm r}$,
which is calculated using available uncertainties from the SDSS and NED,
is ${\rm \sim 20~km~s^{-1}}$.
Pairs of our sample consist of galaxies of comparable luminosities with
mean difference of magnitudes ${\rm \sim 2 ~ mag}$.

In Col.~17 we present BPT \citep[][]{baldwin81} classification types for 27 hosts of SNe presented in SDSS.
The pipeline and the way they were measured is described in \citet[][]{brinchmann04}.
For statistical study, we visually inspected whether the nucleus of the galaxy
is located within the fiber used to obtain its spectrum and
kept only those containing the spectrum of nuclei.
We assign to galaxies their SDSS BPT types using numerical codes:
(-1) for passive nuclei;
(1) for star-forming nuclei;
(3) for composite nuclei (data for MRK 171B composite nuclei were added from \citetalias{H12}), and
(5) for narrow-line Active Galactic Nuclei (AGN).
This coding sequence is used to reflect an increasing sequence of equivalent widths of emission-lines.

\begin{deluxetable}{lllrllrrrrrrrrrrr}
\rotate
\tabletypesize{\scriptsize}
\tablecaption{Data for SNe, their hosts and host neighbors.\label{listofSn}}
\tablewidth{0pt}
\tablehead{
\colhead{SN} & \colhead{SN} & \colhead{$R _{\rm SN}/R _{25}$} & \colhead{SN}
& \colhead{Host name} & \colhead{Neig. name} & \colhead{Host}& \colhead{Host}
& \colhead{Neig.} & \colhead{Neig.} & \colhead{Host}& \colhead{Neig.}
& \colhead{Host} & \colhead{Neig.} & \colhead{$\Delta v_{\rm r}$}& \colhead{$d_{\rm p}$}
& \colhead{Host}\\
\colhead{name} & \colhead{type} & \colhead{} & \colhead{PA}
& \colhead{} & \colhead{} & \colhead{\emph{t}-type}& \colhead{bar}
& \colhead{\emph{t}-type} & \colhead{bar} & \colhead{$mag$}& \colhead{$mag$}
& \colhead{diam.} & \colhead{diam.} & \colhead{}& \colhead{}
& \colhead{BPT}\\
\colhead{}
&\colhead{}
&\colhead{}
&\colhead{$\rm{deg}$}
&\colhead{}
&\colhead{}
&\colhead{}
&\colhead{}
&\colhead{}
&\colhead{}
&\colhead{blue}
&\colhead{blue}
&\colhead{kpc}
&\colhead{kpc}
&\colhead{$\rm{km~s^{-1}}$}
&\colhead{kpc}
&\colhead{}\\
\colhead{(1)}&\colhead{(2)}&\colhead{(3)}&\colhead{(4)}&\colhead{(5)}&\colhead{(6)}&\colhead{(7)}
&\colhead{(8)}&\colhead{(9)}&\colhead{(10)}&\colhead{(11)}&\colhead{(12)}&\colhead{(13)}&\colhead{(14)}
&\colhead{(15)}&\colhead{(16)}&\colhead{(17)}
}
\startdata
1960I&Ia:&0.42&65&MRK 210&J122702.87+481756.4&3&1&0&0&-20.3&-18.3&30.7&10.6&111&54.1& \\
1960J&&0.96&107&NGC 4375&J122458.12+283501.5&2&0&2&0&-22.3&-19.4&51.1&16.0&222&57.7&5 \\
1960M&I&0.67&11&MRK 386&KUG 0816+221A&4&1&4&0&-20.9&-18.6&26.2&8.9&118&25.1&\\
1963D&Ia*&0.74&162&NGC 4146&J121009.97+262527.7  &3&1&1&0&-21.1&-17.0&39.0&7.0&200&49.9&-1\\
1963K&I:&0.31&124&NGC 3656 &SBS 1120+540&0&0&0&0&-20.1&-17.0&28.8&8.4&78&57.9&3\\
1973C&&0.18&41&NGC 3656 &SBS 1120+540&0&0&0&0&-20.1&-17.0&28.8&8.4&78&57.9&3\\
1970L&Ia*&1.83&33&NGC 2968 &MRK 405&-2&1&-2&0&-19.3&-17.3&17.8&6.2&93&34.4&-1\\
1987F&IIn&1.00&154&NGC 4615&MRK 780&5&0&-2&0&-21.0&-18.8&31.9&9.5&209&42.0&\\
1988L&Ib&0.21&163&NGC 5480&SBS 1403+509&5&0&10&0&-19.7&-15.4&21.0&3.4&24&60.8&1\\
1989A&Ia&0.51&175&MRK 736&J112804.17+293038.4&4&1&-5&0&-18.0&-16.3&18.2&3.4&286&7.9&\\
\enddata
\tablecomments{Only 10 entries are shown.
Full Table~\ref{listofSn}
is only available at the CDS via anonymous ftp to
\texttt{cdsarc.u-strasbg.fr (130.79.128.5)}.
A portion is shown here for guidance regarding its form and content}
\end{deluxetable}

\section{Statistics and discussions}
\label{statistics}

\subsection{Multivariate factor analysis}
\label{MFA}

\begin{table}[t]
\caption{Varimax rotated normalized orthogonal factor loadings.
Data for 46 SNe sample.\label{MFAfor46}}
\begin{center}
\begin{tabular}{lrrr}
\hline
\hline
Variable & \multicolumn{1}{c}{$F_1$} & \multicolumn{1}{c}{$F_2$} & \multicolumn{1}{c}{$F_3$}\\
\hline
SN type & 0.16&\bf-0.71& 0.00\\
$ R_{\rm SN} / R_{25}$& 0.33& 0.40& 0.35\\
SN PA & 0.03& 0.06& -0.26 \\
Host \emph{t}-type& 0.47&-0.46& 0.43\\
Host bar&0.14& -0.02 &\bf 0.83 \\
Host abs. mag &\bf -0.56 & 0.16 & \bf 0.50\\
Neig. \emph{t}-type& 0.34&\bf -0.70& 0.21 \\
Neig. bar&\bf 0.71& 0.00& 0.10\\
Neig. abs. mag&\bf -0.76& -0.05& 0.37\\
$\Delta v_{\rm r}$  & 0.18&\bf 0.72& -0.05\\
$ d_{\rm p}$& \bf -0.67&0.14& -0.25\\
\hline
Accum. variance& 21~\% & 39~\% & 53~\%\\
\hline
\end{tabular}
\end{center}
\end{table}

The statistical research was conducted in two steps.
First, we applied an exploratory multivariate factor analysis (MFA, e.g., \citealt[][]{tabachnick06})
to look for correlations between all parameters describing SNe,
their hosts and host neighbors, which are collected in Table~\ref{listofSn}.
This statistical  method is similar to the more commonly used principal component analysis (PCA).
The MFA describes the interdependence and grouping patterns of variables in terms of  factors.
Factor loadings are measures of involvement of variables in factor patterns
and can be interpreted like correlation coefficients.
The square of the loading is the variation that a variable has in common with the factor pattern.
The percent of total variance carried by a factor is the mean of squared loadings for a factor.
Applying MFA, we chose as initial variables all the parameters of Table~\ref{listofSn},
assigning to SN types following coding: 1 to Ia, 2 to II, and 3 to Ibc.
This order is chosen not to shade the suggested age sequence for SN progenitors
from older to younger \citep[e.g.,][]{livio01,heger03,eldridge04}.
To increase confidence on the obtained results, analysis was conducted for two data sets.
First, we run MFA for all initial variables of Table~\ref{listofSn}, excluding SN host BPT classes only.
In this case, the total number of SNe in the statistics is 46 in 40 hosts.
Then, we run MFA including also host BPT classes as the initial variable.
In this case, the total number of SNe in the statistics is reduced to 21 in 16 hosts (see Table~\ref{listofSn}).
In order to simplify the interpretation of the results,
we only present the rotated varimax normalized orthogonal values for the three most significant factors,
with highlighted values above $0.5$ correlation threshold.
Table~\ref{MFAfor46} shows the factor loadings, i.e., the correlation coefficients
between the initial variables and the factors for the $N = 46$ SNe sample
with all initial variables of Table~\ref{listofSn} excluding BPT classes of hosts.
Table~\ref{MFAfor21} shows factor loadings for the $N = 21$ SNe sample
with all initial variables of Table~\ref{listofSn} included\footnote{\footnotesize
{MFA coefficients are invariant with respect to linear transformations of any initial parameter,
so our BPT coding does not affect the MFA results significantly.}}.

\begin{table}[t]
\caption{Varimax rotated normalized orthogonal factor loadings.
Data for 21 SNe sample.\label{MFAfor21}}
\begin{center}
\begin{tabular}{lrrr}
\hline
\hline
Variable & \multicolumn{1}{c}{$F_1$} & \multicolumn{1}{c}{$F_2$} & \multicolumn{1}{c}{$F_3$}\\
\hline
SN type &-0.18&\bf 0.84 &0.18\\
$ R_{\rm SN} / R_{25}$&\bf 0.70 &-0.30& -0.13\\
SN PA &0.03 &0.08& -0.42 \\
Host \emph{t}-type&\bf 0.54&\bf 0.74& 0.08\\
Host bar&\bf 0.72& 0.09& -0.09 \\
Host abs. mag & -0.19& -0.15&\bf -0.76\\
Neig. \emph{t}-type& 0.27&\bf 0.83& 0.01 \\
Neig. bar&\bf 0.67& 0.26 &0.46\\
Neig. abs. mag&\bf-0.58& -0.12& \bf -0.66\\
$\Delta v_{\rm r}$  & 0.07 &\bf-0.61& \bf 0.58\\
$ d_{\rm p}  $ &\bf -0.73 &-0.45& -0.22\\
Host BPT  type & -0.17& 0.20 &\bf0.83\\
\hline
Accum. variance& 23~\% & 46~\% & 67~\%\\
\hline
\end{tabular}
\end{center}
\end{table}

In Table~\ref{MFAfor46}, the first factor ($F_1$),
which accounts for about 21~\% of the common variance,
is the combination of the hosts and its neighbors absolute magnitudes,
existence of bar of neighbor and linear projected distance between galaxies in pairs.
Galaxies located in closer pairs have higher luminosities with preferable existence of bar.
Factor $F_2$, which accounts for about 18~\% of the common variance,
is the combination of velocity difference between galaxies in pairs,
host and its neighbor morphologies and SN types.
Members of pairs with smaller velocity differences are galaxies of later morphological types,
and discovered SNe in these pairs are preferably CC with higher rates of Ibc SNe.
Factor $F_3$, which accounts for about 14~\% of the common variance,
is the combination of host absolute magnitudes and the existence of a bar.

In Table~\ref{MFAfor21}, the first factor ($F_1$),
which accounts for about 23~\% of the common variance,
is the combination of SN relative distance from host nuclei, hosts
morphological types and existence of bar, as well as linear projected distance between galaxies in pairs.
Closer located galaxies in pairs have relatively later morphological types and prominent bars.
In such pairs, SNe in their hosts are discovered at larger distances from the nuclei.
Factor $F_2$, which accounts for about 23~\% of the common variance,
corresponds to the factor  $F_2$ of the first MFA and combines the same parameters.
Factor $F_3$, which accounts for about 21~\% of the common variance,
combines  BPT classes of host galaxies, hosts and neighbors absolute magnitudes
and velocity difference between galaxies in pairs.

Combining the results of both MFA analyses, presented in Tables~\ref{MFAfor46} and \ref{MFAfor21},
first we can conclude that change of number of SNe as well as change of used initial variable sets
does not significantly affect most of the obtained results.
Most conclusions repeat each other in different factors of these MFA analyses.
Summarizing the results, we can make the following conclusions.
Nuclear  activity level correlates with host galaxy luminosity.
This is a well known fact proven with many observations \cite[e.g.,][]{meurs84}.
More luminous galaxies located in closer pairs have a higher percentage of bars,
probably triggered by interactions \cite[e.g.,][]{mendezabreu12}.
MFA does not show any correlation between PA and any physical parameters of the pair and the SN.
Comparing the distribution of SNe by angle describing their orientation
with respect to the neighbor galaxy, we found that it is statistically the same as an isotropic distribution,
and is independent of the $\Delta v_{\rm r}$ and  $d_{\rm p}$ of the pair.
This result is in agreement with \cite{petrosian95} and \cite{navasardyan01},
and it can be explained by different distributions of
dynamical timescales of interacting/merging galaxies in pairs.
Another important result of MFA is that the difference of radial velocities
correlates well with type of SNe located in these pairs, this result is discussed below.

\subsection{Morphologies of hosts and neighbors}
\label{morph}

We study the most prominent grouped parameters in detail by exploring one-to-one correlations.
The first row of Table~\ref{morphs} shows mean morphological \emph{t}-types
of our paired sample hosts for SNe of different types.
The errors correspond to the standard error of the mean.
It is obvious that hosts of type Ibc and II SNe are of later morphological classes
than those of type Ia in accordance with MFA.
This is well known observational result \cite[e.g.,][]{cappellaro99,vandenbergh05}.
For comparison, in the second row of Table~\ref{morphs}, mean morphologies of the unbiased sample of
1021 nearby (${\rm \leq 100 ~Mpc}$) hosts of different types SNe from \citetalias{H12} are presented.
As is seen from the results of the Kolmogorov-Smirnov (KS) test for paired hosts of our sample and from sample of \citetalias{H12},
there is no statistically significant difference between our and \citetalias{H12} \emph{t}-types.
In agreement with MFA, SN hosts of our sample tend to form pairs with neighbors
(data of which are presented in the third row of Table~\ref{morphs}) with similar morphologies.
However, if we compare the distributions of \emph{t}-types of all hosts
and all neighbors  irrespectively of SN types,
a KS test shows a statistically significant difference between them.
This probably reflects the fact that the KS test rarely rejects the null hypothesis in small samples.
According to Table~\ref{morphs}, the percentage of barred galaxies among our hosts is larger ($\sim2\sigma$)
than that in the general sample of hosts from \citetalias{H12}.
Since \citetalias{H12} general sample contains both paired and isolated hosts,
but our sample consists of paired hosts only,
the excess of bars in our sample is expected \cite[e.g.,][]{mendezabreu12}.
\begin{table*}[t]
\caption{Morphological classification of SN hosts and their neighbors.\label{morphs}}
\begin{center}
\begin{tabular}{llr@{$\pm$}lr@{$\pm$}lr@{$\pm$}lr@{$\pm$}l}
\tableline
\tableline
&&\multicolumn{2}{c}{All SNe}&\multicolumn{2}{c}{Ia}&\multicolumn{2}{c}{Ibc}&\multicolumn{2}{c}{II}\\
\tableline
1 & Our host mean \emph{t}-type & 3.4 & 0.4 & 2.0 & 0.6 & 4.2 & 0.7 & 5.3 & 0.7 \\
2 & \citetalias{H12} host sample, mean \emph{t}-type & 3.9 & 0.1 & 2.3 & 0.2 & 4.5 & 0.2 & 5.0 & 0.1 \\
3 & Our neighbor mean \emph{t}-type & 4.1 & 0.5 & 2.5 & 0.5 & 5.9 & 0.9 & 4.7 & 1.1 \\
4 & \emph{p} value KS test for rows 1 \& 2 & \multicolumn{2}{c}{0.43} & \multicolumn{2}{c}{0.74} & \multicolumn{2}{c}{0.43} & \multicolumn{2}{c}{0.89} \\	
5 & \emph{p} value KS test for rows 1 \& 3 & \multicolumn{2}{c}{0.00} & \multicolumn{2}{c}{0.14} & \multicolumn{2}{c}{0.71} & \multicolumn{2}{c}{0.46} \\
6 & Our host bar (in \%) & 45 & ~7 & 42 & 12 & 50 & 15 & 47 & 13 \\
7 & \citetalias{H12} host bar (in \%) & 30 & ~1 & 27 & ~3 & 28 & ~4 & 31 & ~2 \\
8 & Our neighbor bar (in \%) & 32 & ~6 & 26 & 10 & 42 & 15 & 33 & 13 \\
\tableline
\end{tabular}
\end{center}
\end{table*}
\begin{figure}[h]
\begin{center}
\includegraphics[width=0.74\hsize,angle=-90]{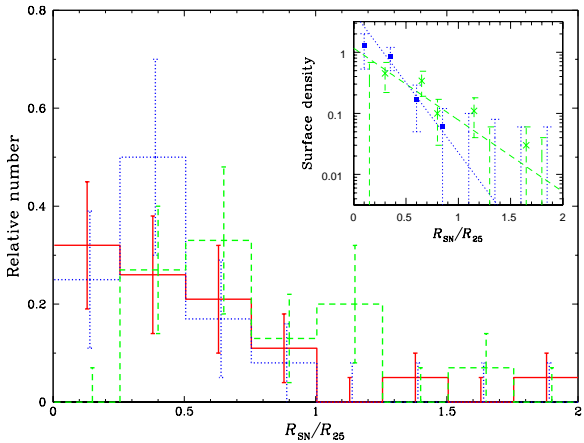}
\caption{Normalized histograms of radial distributions of types Ia (\emph{red solid}),
Ibc (\emph{blue dotted}), and II (\emph{green dashed}) SNe. The \emph{error bars} assume a Poisson distribution (with $\pm1$ object
if none is found). \emph{Top-right}: surface density profiles
(with arbitrary normalization) of all CC SNe:
Ibc (\emph{blue dotted}),  and II (\emph{green dashed}).
The \emph{lines} show the maximum likelihood exponential surface density profiles of CC SNe.
The observational deficit of SNe in the central regions is due to the \cite{shaw79} effect}
\label{radDistr}
\end{center}
\end{figure}

\subsection{Radial distributions}
\label{radial}

The radial distribution of SNe is shown in Fig.~\ref{radDistr}.
The mean value of $R_{\rm SN}/R_{25}$ is $0.53\pm0.10$ for type Ia.
The mean normalized distance $R_{\rm SN}/R_{25}$ of SNe II, at $0.74\pm0.09$
is roughly double that of SNe Ibc at $0.38\pm0.06$. The significance of more concentrated Ibc SNe relative to II SNe is $3.35\sigma$.
The mean values of normalized distances of Ia SNe agree with
the results of Fig.~4 in \cite{vandenbergh97} calculated in the same way without deprojecting.
Our mean values for CC SNe within estimated errors are in agreement with those of \citetalias{H09},
where values $0.62\pm0.03$ for type II and $0.45\pm0.04$ for type Ibc SNe were reported.
The higher II to Ibc ratio of mean normalized distances in our sample
is marginally significant ($1.54\sigma$) than that in the \citetalias{H09} general larger sample.

\begin{table*}[t]
\caption{The mean values of parameters of pairs containing SNe of different types.\label{pairProperties}}
\begin{center}
\begin{tabular}{lcr@{$\pm$}lr@{$\pm$}l}
\tableline
\tableline
SN type & Number of SNe & \multicolumn{2}{c}{Mean $\Delta v_{\rm r}~{\rm (km~s^{-1})}$} & \multicolumn{2}{c}{Mean $d_{\rm p}~{\rm (kpc)}$} \\
\tableline
Ia & 19 & 134 & 27 & 34 & 4 \\
Ibc & 12 & 56 & 14 & 28 & 5 \\
II & 15 & 129 & 18 & 32 & 6 \\
All (including unclassified) & 56 & 117 & ~9 & 33 & 3 \\
\tableline
\end{tabular}
\end{center}
\end{table*}

In addition, we calculated surface densities of type Ibc and II SNe,
assuming that they are located within discs of spiral hosts, as is shown in top-right corner of Fig.~\ref{radDistr}.
\cite{wang10} measured number and surface density distributions of type II SNe in their hosts
and indicated that SNe detected in SF hosts follow an exponential law,
but the distribution of type II SNe in AGN hosts (which are in general thought to be more disturbed)
significantly deviates from exponential law.
Moreover, \cite{habergham12} found a remarkable excess of Ibc SNe within central regions of disturbed galaxies,
i.e. mainly in galaxies showing signs of merger-triggered starbursts in their nuclei.
To check consistency of surface density distributions of CC SNe with an exponential model,
we generated exponential distributions with the corresponding scale lengths using maximum likelihood fitting
according to \citetalias{H09} and compared real SNe distributions with them by means of a KS test.
We found a deviation from the exponential density fall for SNe Ibc.
To figure out  the nature of the deviation, we removed all the SNe within $R_{\rm SN}/R_{25}<0.25$ bin,
generated new exponential distributions and performed a KS test again. The deviation vanished.
Therefore, the SNe located closer to the nuclei are responsible for the deviation from exponential distribution.
The surface densities shown in Fig.~\ref{radDistr} suggest that
the observational deficit of SNe in the central regions is due to the \cite{shaw79} effect as the main source of the deviation.

There is non-significant excess of Ibc SNe in central regions of hosts of our sample
in comparison with those of \citetalias{H09} sample.
This result, not reproducing that of \cite{habergham12},
can be explained by large number of disturbed galaxies in our sample,
hence less accurate measurements of $R_{25}$
and less secure inclination corrections of SN distances.

We also checked how kinematical properties of pairs,
such as $\Delta v_{\rm r}$ and $d_{\rm p}$, describing strength/stage of interacting/merging,
affect the distribution of SNe within their hosts.
We divided our pairs sample for each SN type into two subsamples,
with larger and smaller  $\Delta v_{\rm r}$, and we checked whether SNe in pairs with small $\Delta v_{\rm r}$
have different radial distances from centers of their hosts.
For CC SNe, we found this difference insignificant.
For Ia SNe, there is no difference at all.
The similar procedure was done for $d_{\rm p}$ with the same results.

\subsection{Pair properties}
\label{pairs}

In addition, we analyzed subsamples of pairs with SNe of different types,
with the main results presented in Table~\ref{pairProperties}.
A KS test shows that the distribution of $\Delta v_{\rm r}$ of pairs containing SNe Ia and II
is the same with practically consistent mean values.
In contrast, the same distributions of pairs with Ibc and II SNe (also of pairs with SNe Ibc and non-Ibc)
are significantly different ($3.2\sigma$) with smaller mean $\Delta v_{\rm r}$ for Ibc.
This means that Ibc SNe explode preferably in pairs with stronger interaction \citep[e.g.,][]{patton00}.
We consider this as an important result needing a physical interpretation.
The same results were qualitatively obtained using MFA.
It is worth noting, that there is no significant difference between mean values of $d_{\rm p}$
of pairs with SNe of different types as is seen in Table~\ref{pairProperties}.
Assuming that sizes of galaxies can bias dependence of pair properties on $d_{\rm p}$,
we also normalized projected separations to sizes of pair members ($d_{\rm p}/\Sigma{R_{25}}$).
This normalization does not add any statistical change to all the correlations with $d_{\rm p}$.
Since distance may be an important factor biasing SN types and host properties of our sample,
we also compared mean distances of hosts of different type SNe
and did not find any differences between them.

To explain the strong dependence of CC SNe types on pair properties,
we considered SFR as the main parameter, which can affect SN production in galaxies.
Because of the small sample of hosts with measured SFR from SDSS spectra,
we do not have enough data for making any significant direct conclusion.
Therefore, we can only make interpretative suggestions based on other large sample studies.
In general, SN rate and number ratio of type Ibc to type II SNe
increases with increasing of SFR \cite[e.g.,][]{cappellaro99,mannucci05, petrosian05, boissier09}.
In this respect, we expect a relatively larger amount of star-forming galaxies,
especially with smaller $\Delta v_{\rm r}$  and $d_{\rm p}$,
in our sample of paired hosts due to interaction-triggered starbursts.
Therefore, excess of Ibc SNe compared to II SNe in the pairs with smaller $\Delta v_{\rm r}$
and $d_{\rm p}$ can be a result of higher SFR.
In close environments of galaxies ($d_{\rm p} < ~60~{\rm kpc}$)
dependence of SFR indicators on $\Delta v_{\rm r}$ is stronger than
that on $d_{\rm p}$ as was shown in \cite{barton00,ellison08,patton11}.
The fact, that number ratio of SN types depends on $\Delta v_{\rm r}$
much more strongly than on $d_{\rm p}$, is supported by these results.
We suggest the cut-off by larger value of $d_{\rm p}$ as a possible improvement
to make the correlation between SN type and $d_{\rm p}$ more apparent.

Since mass ratios of pair components also can influence SFR \cite[e.g.,][]{kennicutt98,cox08},
we analyzed pairs with different SN types to search whether
there is a difference between luminosity ratios of pair components
(to see whether Ibc SNe occur preferably in pairs of major mergers).
We found that there is no statistical difference between these pairs.

\section{Summary}
\label{summary}

We have studied 56 SNe located in pairs of galaxies and analyzed
the dependence of SN properties on the host and neighbor properties.
The main results are:

\begin{enumerate}
\item
SN hosts as well as their neighbors in our sample have statistically the same morphological classification
as hosts from the general sample of SNe ($\leq ~100~{\rm Mpc}$) from \citetalias{H12}.
Our sample contains only paired galaxies, therefore the amount of barred hosts in our sample
is higher than in the general sample of SN hosts.
\item
Mean radial distances obtained for our sample of SN hosts
are in agreement with those reported in \cite{vandenbergh97} and \citetalias{H09}.
The ratio of mean radial distances of type II and Ibc SNe is about 2.
Radial distributions of CC SNe in our sample of paired galaxies
are difficult to fit into a model with exponentially falling surface density
because of the observational deficit of SNe in regions located close to nuclei.

\item
SNe of type Ibc are located in pairs having smaller $\Delta v_{\rm r}$
than pairs containing type II and Ia SNe. This difference is significant at the $3.2\sigma$ level.
We suggest that this result is because of higher SFR in closer pairs,
resulting in a higher rate of Ibc SNe compared to II SNe.
A similar trend is found considering  $d_{\rm p}$ of pairs,
although it is significantly weaker.
\item
The luminosity ratio of galaxies in pairs does not display any correlation with the SN types.
\item
The orientation of SNe with respect to the preferred direction is found to be  isotropic,
independent of $\Delta v_{\rm r}$ and $d_{p}$.
\end{enumerate}

As a conclusion, we consider that close environment of galaxies
can have some observable effect on SN production
due to the impact on SF of galaxies.

\footnotesize
\acknowledgments
A.R.P, A.A.H., and L.S.A. acknowledge the hospitality of the Institut d'Astrophysique de Paris (France)
during their stay as visiting scientists supported by
the Collaborative Bilateral Research Project of the State Committee of Science (SCS) of the Republic of Armenia
and the French Centre National de la Recherch\'e Scientifique (CNRS).
This work was made possible in part by a research grant from the
Armenian National Science and Education Fund (ANSEF) based in New York, USA.
V.Zh.A. is supported by grant SFRH/BPD/70574/2010 from FCT (Portugal)
and would further like to thank for the support by the ERC
under the FP7/EC through a Starting Grant agreement number 239953. We gratefully
acknowledge the Anonymous Referee for the constructive comments and
suggestions.
This research made use of the NASA/IPAC Extragalactic Database (NED),
which is available at \texttt{http://ned.ipac.caltech.edu/},
and operated by the Jet Propulsion Laboratory,
California Institute of Technology,
under contract with the National Aeronautics and Space Administration.
Funding for SDSS-III has been provided by the Alfred P.~Sloan Foundation,
the Participating Institutions, the National Science Foundation,
and the US Department of Energy Office of Science.
The SDSS-III web site is \texttt{http://www.sdss3.org/}.
SDSS-III is managed by the Astrophysical
Research Consortium for the Participating Institutions of the SDSS-III Collaboration including
the University of Arizona, the Brazilian Participation Group, Brookhaven National Laboratory,
University of Cambridge, Carnegie Mellon University, University of Florida, the French Participation Group,
the German Participation Group, Harvard University, the Instituto de Astrofisica de Canarias,
the Michigan State/Notre Dame/JINA Participation Group, Johns Hopkins University,
Lawrence Berkeley National Laboratory, Max Planck Institute for Astrophysics,
Max Planck Institute for Extraterrestrial Physics, New Mexico State University, New York University,
Ohio State University, Pennsylvania State University, University of Portsmouth, Princeton University,
the Spanish Participation Group, University of Tokyo, University of Utah, Vanderbilt University,
University of Virginia, University of Washington, and Yale University.
This publication makes use of data products from the Digitized Sky Survey produced at
the Space Telescope Science Institute under US Government grant NAG W-2166.
The images of this survey are based on photographic data obtained using the
Oschin Schmidt Telescope on Palomar Mountain and the UK Schmidt Telescope.
The plates were processed into the present digital form with the permission of these institutions.
The Second Palomar Observatory Sky Survey (POSS-II) was made by the
California Institute of Technology with funds from the National Science Foundation,
the National Aeronautics and Space Administration,
the National Geographic Society, the Sloan Foundation,
the Samuel Oschin Foundation, and the Eastman Kodak Corporation.


\begin{thebibliography}{60}
\ifx \bisbn   \undefined \def \bisbn  #1{ISBN #1}\fi
\ifx \binits  \undefined \def \binits#1{#1} \fi
\ifx \bauthor  \undefined \def \bauthor#1{#1} \fi
\ifx \batitle  \undefined \def \batitle#1{#1} \fi
\ifx \bjtitle  \undefined \def \bjtitle#1{#1}\fi
\ifx \bvolume  \undefined \def \bvolume#1{\textbf{#1}}\fi
\ifx \byear  \undefined \def \byear#1{#1} \fi
\ifx \bissue  \undefined \def \bissue#1{#1} \fi
\ifx \bfpage  \undefined \def \bfpage#1{#1} \fi
\ifx \blpage  \undefined \def \blpage #1{#1} \fi
\ifx \burl  \undefined \def \burl#1{\textsf{#1}} \fi
\ifx \doiurl  \undefined \def \doiurl#1{\textsf{#1}} \fi
\ifx \betal  \undefined \def \betal{\textit{et al.}} \fi
\ifx \binstitute  \undefined \def \binstitute#1{#1} \fi
\ifx \binstitutionaled  \undefined \def \binstitutionaled#1{#1} \fi
\ifx \bctitle  \undefined \def \bctitle#1{#1} \fi
\ifx \beditor  \undefined \def \beditor#1{#1} \fi
\ifx \bpublisher  \undefined \def \bpublisher#1{#1} \fi
\ifx \bbtitle  \undefined \def \bbtitle#1{#1} \fi
\ifx \bedition  \undefined \def \bedition#1{#1} \fi
\ifx \bseriesno  \undefined \def \bseriesno#1{#1} \fi
\ifx \blocation  \undefined \def \blocation#1{#1} \fi
\ifx \bsertitle  \undefined \def \bsertitle#1{#1} \fi
\ifx \bsnm \undefined \def \bsnm#1{#1} \fi
\ifx \bsuffix \undefined \def \bsuffix#1{#1} \fi
\ifx \bparticle \undefined \def \bparticle#1{#1} \fi
\ifx \barticle \undefined \def \barticle#1{#1} \fi
\ifx \bconfdate \undefined \def \bconfdate #1{#1} \fi
\ifx \botherref \undefined \def \botherref #1{#1} \fi
\ifx \url \undefined \def \url#1{\textsf{#1}} \fi
\ifx \bchapter \undefined \def \bchapter#1{#1} \fi
\ifx \bbook \undefined \def \bbook#1{#1} \fi
\ifx \bcomment \undefined \def \bcomment#1{#1} \fi
\ifx \oauthor \undefined \def \oauthor#1{#1} \fi
\ifx \citeauthoryear \undefined \def \citeauthoryear#1{#1} \fi
\ifx \endbibitem  \undefined \def \endbibitem {}\fi
\ifx \bconflocation  \undefined \def \bconflocation#1{#1} \fi
\ifx \arxivurl  \undefined \def \arxivurl#1{\textsf{#1}} \fi

\bibitem[\protect\citeauthoryear{{Ahn} et~al.}{2012}]{ahn12}
\begin{barticle}
\bauthor{\bsnm{{Ahn}}, \binits{C.P.}},
\bauthor{\bsnm{{Alexandroff}}, \binits{R.}},
\bauthor{\bsnm{{Allende Prieto}}, \binits{C.}},
\bauthor{\bparticle{et} \bsnm{al.}}:
\bjtitle{\apjs}
\bvolume{203},
\bfpage{21}
(\byear{2012})
\end{barticle}
\endbibitem

\bibitem[\protect\citeauthoryear{{Anderson} and {James}}{2008}]{anderson08}
\begin{barticle}
\bauthor{\bsnm{{Anderson}}, \binits{J.P.}},
\bauthor{\bsnm{{James}}, \binits{P.A.}}:
\bjtitle{\mnras}
\bvolume{390},
\bfpage{1527}
(\byear{2008})
\end{barticle}
\endbibitem

\bibitem[\protect\citeauthoryear{{Aramyan} et~al.}{2013}]{aramyan13}
\begin{barticle}
\bauthor{\bsnm{{Aramyan}}, \binits{L.S.}},
\bauthor{\bsnm{{Petrosian}}, \binits{A.R.}},
\bauthor{\bsnm{{Hakobyan}}, \binits{A.A.}},
\bauthor{\bparticle{et} \bsnm{al.}}:
\bjtitle{Astrophysics}
\bvolume{56},
\bfpage{153}
(\byear{2013})
\end{barticle}
\endbibitem

\bibitem[\protect\citeauthoryear{{Baldwin} et~al.}{1981}]{baldwin81}
\begin{barticle}
\bauthor{\bsnm{{Baldwin}}, \binits{J.A.}},
\bauthor{\bsnm{{Phillips}}, \binits{M.M.}},
\bauthor{\bsnm{{Terlevich}}, \binits{R.}}:
\bjtitle{\pasp}
\bvolume{93},
\bfpage{5}
(\byear{1981})
\end{barticle}
\endbibitem

\bibitem[\protect\citeauthoryear{{Barton} et~al.}{2000}]{barton00}
\begin{barticle}
\bauthor{\bsnm{{Barton}}, \binits{E.J.}},
\bauthor{\bsnm{{Geller}}, \binits{M.J.}},
\bauthor{\bsnm{{Kenyon}}, \binits{S.J.}}:
\bjtitle{\apj}
\bvolume{530},
\bfpage{660}
(\byear{2000})
\end{barticle}
\endbibitem

\bibitem[\protect\citeauthoryear{{Bartunov} et~al.}{1994}]{bartunov94}
\begin{barticle}
\bauthor{\bsnm{{Bartunov}}, \binits{O.S.}},
\bauthor{\bsnm{{Tsvetkov}}, \binits{D.Y.}},
\bauthor{\bsnm{{Filimonova}}, \binits{I.V.}}:
\bjtitle{\pasp}
\bvolume{106},
\bfpage{1276}
(\byear{1994})
\end{barticle}
\endbibitem

\bibitem[\protect\citeauthoryear{{Blanton} and {Moustakas}}{2009}]{blanton09}
\begin{barticle}
\bauthor{\bsnm{{Blanton}}, \binits{M.R.}},
\bauthor{\bsnm{{Moustakas}}, \binits{J.}}:
\bjtitle{\araa}
\bvolume{47},
\bfpage{159}
(\byear{2009})
\end{barticle}
\endbibitem

\bibitem[\protect\citeauthoryear{{Boissier} and {Prantzos}}{2009}]{boissier09}
\begin{barticle}
\bauthor{\bsnm{{Boissier}}, \binits{S.}},
\bauthor{\bsnm{{Prantzos}}, \binits{N.}}:
\bjtitle{\aap}
\bvolume{503},
\bfpage{137}
(\byear{2009})
\end{barticle}
\endbibitem

\bibitem[\protect\citeauthoryear{{Bottinelli} et~al.}{1995}]{bottinelli95}
\begin{barticle}
\bauthor{\bsnm{{Bottinelli}}, \binits{L.}},
\bauthor{\bsnm{{Gouguenheim}}, \binits{L.}},
\bauthor{\bsnm{{Paturel}}, \binits{G.}},
\bauthor{\bsnm{{Teerikorpi}}, \binits{P.}}:
\bjtitle{\aap}
\bvolume{296},
\bfpage{64}
(\byear{1995})
\end{barticle}
\endbibitem

\bibitem[\protect\citeauthoryear{{Bournaud}}{2011}]{bournaud11}
\begin{bchapter}
\bauthor{\bsnm{{Bournaud}}, \binits{F.}}:
In: \beditor{\bsnm{{Charbonnel}}, \binits{C.}},
\beditor{\bsnm{{Montmerle}}, \binits{T.}} (eds.)
\bsertitle{EAS Publications Series},
vol. \bseriesno{51},
p. \bfpage{107}
(\byear{2011})
\end{bchapter}
\endbibitem

\bibitem[\protect\citeauthoryear{{Bressan} et~al.}{2002}]{bressan02}
\begin{barticle}
\bauthor{\bsnm{{Bressan}}, \binits{A.}},
\bauthor{\bsnm{{Della Valle}}, \binits{M.}},
\bauthor{\bsnm{{Marziani}}, \binits{P.}}:
\bjtitle{\mnras}
\bvolume{331},
\bfpage{25}
(\byear{2002})
\end{barticle}
\endbibitem

\bibitem[\protect\citeauthoryear{{Brinchmann} et~al.}{2004}]{brinchmann04}
\begin{barticle}
\bauthor{\bsnm{{Brinchmann}}, \binits{J.}},
\bauthor{\bsnm{{Charlot}}, \binits{S.}},
\bauthor{\bsnm{{White}}, \binits{S.D.M.}},
\bauthor{\bparticle{et} \bsnm{al.}}:
\bjtitle{\mnras}
\bvolume{351},
\bfpage{1151}
(\byear{2004})
\end{barticle}
\endbibitem

\bibitem[\protect\citeauthoryear{{Cappellaro} et~al.}{1999}]{cappellaro99}
\begin{barticle}
\bauthor{\bsnm{{Cappellaro}}, \binits{E.}},
\bauthor{\bsnm{{Evans}}, \binits{R.}},
\bauthor{\bsnm{{Turatto}}, \binits{M.}}:
\bjtitle{\aap}
\bvolume{351},
\bfpage{459}
(\byear{1999})
\end{barticle}
\endbibitem

\bibitem[\protect\citeauthoryear{{Cox} et~al.}{2008}]{cox08}
\begin{barticle}
\bauthor{\bsnm{{Cox}}, \binits{T.J.}},
\bauthor{\bsnm{{Jonsson}}, \binits{P.}},
\bauthor{\bsnm{{Somerville}}, \binits{R.S.}},
\bauthor{\bparticle{et} \bsnm{al.}}:
\bjtitle{\mnras}
\bvolume{384},
\bfpage{386}
(\byear{2008})
\end{barticle}
\endbibitem

\bibitem[\protect\citeauthoryear{{Di Matteo} et~al.}{2007}]{dimatteo07}
\begin{barticle}
\bauthor{\bsnm{{Di Matteo}}, \binits{P.}},
\bauthor{\bsnm{{Combes}}, \binits{F.}},
\bauthor{\bsnm{{Melchior}}, \binits{A.-L.}},
\bauthor{\bsnm{{Semelin}}, \binits{B.}}:
\bjtitle{\aap}
\bvolume{468},
\bfpage{61}
(\byear{2007})
\end{barticle}
\endbibitem

\bibitem[\protect\citeauthoryear{{Eldridge} and {Tout}}{2004}]{eldridge04}
\begin{barticle}
\bauthor{\bsnm{{Eldridge}}, \binits{J.J.}},
\bauthor{\bsnm{{Tout}}, \binits{C.A.}}:
\bjtitle{\mnras}
\bvolume{353},
\bfpage{87}
(\byear{2004})
\end{barticle}
\endbibitem

\bibitem[\protect\citeauthoryear{{Ellison} et~al.}{2008}]{ellison08}
\begin{barticle}
\bauthor{\bsnm{{Ellison}}, \binits{S.L.}},
\bauthor{\bsnm{{Patton}}, \binits{D.R.}},
\bauthor{\bsnm{{Simard}}, \binits{L.}},
\bauthor{\bsnm{{McConnachie}}, \binits{A.W.}}:
\bjtitle{\aj}
\bvolume{135},
\bfpage{1877}
(\byear{2008})
\end{barticle}
\endbibitem

\bibitem[\protect\citeauthoryear{{Ellison} et~al.}{2011}]{ellison11}
\begin{barticle}
\bauthor{\bsnm{{Ellison}}, \binits{S.L.}},
\bauthor{\bsnm{{Patton}}, \binits{D.R.}},
\bauthor{\bsnm{{Mendel}}, \binits{J.T.}},
\bauthor{\bsnm{{Scudder}}, \binits{J.M.}}:
\bjtitle{\mnras}
\bvolume{418},
\bfpage{2043}
(\byear{2011})
\end{barticle}
\endbibitem

\bibitem[\protect\citeauthoryear{{Gyulzadyan} et~al.}{2011}]{G11}
\begin{barticle}
\bauthor{\bsnm{{Gyulzadyan}}, \binits{M.}},
\bauthor{\bsnm{{McLean}}, \binits{B.}},
\bauthor{\bsnm{{Adibekyan}}, \binits{V.Z.}},
\bauthor{\bparticle{et} \bsnm{al.}}:
\bjtitle{Astrophysics}
\bvolume{54},
\bfpage{15}
(\byear{2011})
\end{barticle}
\endbibitem

\bibitem[\protect\citeauthoryear{{Habergham} et~al.}{2012}]{habergham12}
\begin{barticle}
\bauthor{\bsnm{{Habergham}}, \binits{S.M.}},
\bauthor{\bsnm{{James}}, \binits{P.A.}},
\bauthor{\bsnm{{Anderson}}, \binits{J.P.}}:
\bjtitle{\mnras}
\bvolume{424},
\bfpage{2841}
(\byear{2012})
\end{barticle}
\endbibitem

\bibitem[\protect\citeauthoryear{{Hakobyan}}{2008}]{hakobyan08b}
\begin{barticle}
\bauthor{\bsnm{{Hakobyan}}, \binits{A.A.}}:
\bjtitle{Astrophysics}
\bvolume{51},
\bfpage{69}
(\byear{2008})
\end{barticle}
\endbibitem

\bibitem[\protect\citeauthoryear{{Hakobyan} et~al.}{2008}]{hakobyan08a}
\begin{barticle}
\bauthor{\bsnm{{Hakobyan}}, \binits{A.A.}},
\bauthor{\bsnm{{Petrosian}}, \binits{A.R.}},
\bauthor{\bsnm{{McLean}}, \binits{B.}},
\bauthor{\bparticle{et} \bsnm{al.}}:
\bjtitle{\aap}
\bvolume{488},
\bfpage{523}
(\byear{2008})
\end{barticle}
\endbibitem

\bibitem[\protect\citeauthoryear{{Hakobyan} et~al.}{2009}]{H09}
\begin{barticle}
\bauthor{\bsnm{{Hakobyan}}, \binits{A.A.}},
\bauthor{\bsnm{{Mamon}}, \binits{G.A.}},
\bauthor{\bsnm{{Petrosian}}, \binits{A.R.}},
\bauthor{\bparticle{et} \bsnm{al.}}:
\bjtitle{\aap}
\bvolume{508},
\bfpage{1259}
(\byear{2009})
\end{barticle}
\endbibitem

\bibitem[\protect\citeauthoryear{{Hakobyan} et~al.}{2011}]{hakobyan11}
\begin{barticle}
\bauthor{\bsnm{{Hakobyan}}, \binits{A.A.}},
\bauthor{\bsnm{{Petrosian}}, \binits{A.R.}},
\bauthor{\bsnm{{Mamon}}, \binits{G.A.}},
\bauthor{\bparticle{et} \bsnm{al.}}:
\bjtitle{Astrophysics}
\bvolume{54},
\bfpage{301}
(\byear{2011})
\end{barticle}
\endbibitem

\bibitem[\protect\citeauthoryear{{Hakobyan} et~al.}{2012}]{H12}
\begin{barticle}
\bauthor{\bsnm{{Hakobyan}}, \binits{A.A.}},
\bauthor{\bsnm{{Adibekyan}}, \binits{V.Z.}},
\bauthor{\bsnm{{Aramyan}}, \binits{L.S.}},
\bauthor{\bparticle{et} \bsnm{al.}}:
\bjtitle{\aap}
\bvolume{544},
\bfpage{A81}
(\byear{2012})
\end{barticle}
\endbibitem

\bibitem[\protect\citeauthoryear{{Hamuy}}{2003}]{hamuy03}
\begin{botherref}
\oauthor{\bsnm{{Hamuy}}, \binits{M.}}:
\arxivurl{arXiv:astro-ph/0301006}
(2003)
\end{botherref}
\endbibitem

\bibitem[\protect\citeauthoryear{{Heger} et~al.}{2003}]{heger03}
\begin{barticle}
\bauthor{\bsnm{{Heger}}, \binits{A.}},
\bauthor{\bsnm{{Fryer}}, \binits{C.L.}},
\bauthor{\bsnm{{Woosley}}, \binits{S.E.}},
\bauthor{\bparticle{et} \bsnm{al.}}:
\bjtitle{\apj}
\bvolume{591},
\bfpage{288}
(\byear{2003})
\end{barticle}
\endbibitem

\bibitem[\protect\citeauthoryear{{Ho}}{2008}]{ho08}
\begin{barticle}
\bauthor{\bsnm{{Ho}}, \binits{L.C.}}:
\bjtitle{\araa}
\bvolume{46},
\bfpage{475}
(\byear{2008})
\end{barticle}
\endbibitem

\bibitem[\protect\citeauthoryear{{Huchra} et~al.}{1990}]{huchra90}
\begin{barticle}
\bauthor{\bsnm{{Huchra}}, \binits{J.P.}},
\bauthor{\bsnm{{Geller}}, \binits{M.J.}},
\bauthor{\bsnm{{de Lapparent}}, \binits{V.}},
\bauthor{\bsnm{{Corwin}}, \binits{H.G.} \bsuffix{Jr.}}:
\bjtitle{\apjs}
\bvolume{72},
\bfpage{433}
(\byear{1990})
\end{barticle}
\endbibitem

\bibitem[\protect\citeauthoryear{{Jarrett} et~al.}{2006}]{jarrett06}
\begin{barticle}
\bauthor{\bsnm{{Jarrett}}, \binits{T.H.}},
\bauthor{\bsnm{{Polletta}}, \binits{M.}},
\bauthor{\bsnm{{Fournon}}, \binits{I.P.}},
\bauthor{\bparticle{et} \bsnm{al.}}:
\bjtitle{\aj}
\bvolume{131},
\bfpage{261}
(\byear{2006})
\end{barticle}
\endbibitem

\bibitem[\protect\citeauthoryear{{Kennicutt}}{1998}]{kennicutt98}
\begin{barticle}
\bauthor{\bsnm{{Kennicutt}}, \binits{R.C.} \bsuffix{Jr.}}:
\bjtitle{\araa}
\bvolume{36},
\bfpage{189}
(\byear{1998})
\end{barticle}
\endbibitem

\bibitem[\protect\citeauthoryear{{Liu} et~al.}{2012}]{liu12}
\begin{barticle}
\bauthor{\bsnm{{Liu}}, \binits{X.}},
\bauthor{\bsnm{{Shen}}, \binits{Y.}},
\bauthor{\bsnm{{Strauss}}, \binits{M.A.}}:
\bjtitle{\apj}
\bvolume{745},
\bfpage{94}
(\byear{2012})
\end{barticle}
\endbibitem

\bibitem[\protect\citeauthoryear{{Livio}}{2001}]{livio01}
\begin{bchapter}
\bauthor{\bsnm{{Livio}}, \binits{M.}}:
In: \beditor{\bsnm{{Livio}}, \binits{M.}},
\beditor{\bsnm{{Panagia}}, \binits{N.}},
\beditor{\bsnm{{Sahu}}, \binits{K.}} (eds.)
\bbtitle{Supernovae and Gamma-Ray Bursts: the Greatest Explosions since the Big
  Bang},
p. \bfpage{334}
(\byear{2001})
\end{bchapter}
\endbibitem

\bibitem[\protect\citeauthoryear{{Mannucci} et~al.}{2006}]{mannucci06}
\begin{barticle}
\bauthor{\bsnm{{Mannucci}}, \binits{F.}},
\bauthor{\bsnm{{Della Valle}}, \binits{M.}},
\bauthor{\bsnm{{Panagia}}, \binits{N.}}:
\bjtitle{\mnras}
\bvolume{370},
\bfpage{773}
(\byear{2006})
\end{barticle}
\endbibitem

\bibitem[\protect\citeauthoryear{{Mannucci} et~al.}{2005}]{mannucci05}
\begin{barticle}
\bauthor{\bsnm{{Mannucci}}, \binits{F.}},
\bauthor{\bsnm{{Della Valle}}, \binits{M.}},
\bauthor{\bsnm{{Panagia}}, \binits{N.}},
\bauthor{\bparticle{et} \bsnm{al.}}:
\bjtitle{\aap}
\bvolume{433},
\bfpage{807}
(\byear{2005})
\end{barticle}
\endbibitem

\bibitem[\protect\citeauthoryear{{Maoz} and {Mannucci}}{2012}]{maoz12}
\begin{barticle}
\bauthor{\bsnm{{Maoz}}, \binits{D.}},
\bauthor{\bsnm{{Mannucci}}, \binits{F.}}:
\bjtitle{\pasa}
\bvolume{29},
\bfpage{447}
(\byear{2012})
\end{barticle}
\endbibitem

\bibitem[\protect\citeauthoryear{{Maza} and {van den Bergh}}{1976}]{maza76}
\begin{barticle}
\bauthor{\bsnm{{Maza}}, \binits{J.}},
\bauthor{\bsnm{{van den Bergh}}, \binits{S.}}:
\bjtitle{\apj}
\bvolume{204},
\bfpage{519}
(\byear{1976})
\end{barticle}
\endbibitem

\bibitem[\protect\citeauthoryear{{M{\'e}ndez-Abreu}
  et~al.}{2012}]{mendezabreu12}
\begin{barticle}
\bauthor{\bsnm{{M{\'e}ndez-Abreu}}, \binits{J.}},
\bauthor{\bsnm{{S{\'a}nchez-Janssen}}, \binits{R.}},
\bauthor{\bsnm{{Aguerri}}, \binits{J.A.L.}},
\bauthor{\bparticle{et} \bsnm{al.}}:
\bjtitle{\apjl}
\bvolume{761},
\bfpage{6}
(\byear{2012})
\end{barticle}
\endbibitem

\bibitem[\protect\citeauthoryear{{Meurs} and {Wilson}}{1984}]{meurs84}
\begin{barticle}
\bauthor{\bsnm{{Meurs}}, \binits{E.J.A.}},
\bauthor{\bsnm{{Wilson}}, \binits{A.S.}}:
\bjtitle{\aap}
\bvolume{136},
\bfpage{206}
(\byear{1984})
\end{barticle}
\endbibitem

\bibitem[\protect\citeauthoryear{{Mihos} and {Hernquist}}{1996}]{mihos96}
\begin{barticle}
\bauthor{\bsnm{{Mihos}}, \binits{J.C.}},
\bauthor{\bsnm{{Hernquist}}, \binits{L.}}:
\bjtitle{\apj}
\bvolume{464},
\bfpage{641}
(\byear{1996})
\end{barticle}
\endbibitem

\bibitem[\protect\citeauthoryear{{Navasardyan} et~al.}{2001}]{navasardyan01}
\begin{barticle}
\bauthor{\bsnm{{Navasardyan}}, \binits{H.}},
\bauthor{\bsnm{{Petrosian}}, \binits{A.R.}},
\bauthor{\bsnm{{Turatto}}, \binits{M.}},
\bauthor{\bparticle{et} \bsnm{al.}}:
\bjtitle{\mnras}
\bvolume{328},
\bfpage{1181}
(\byear{2001})
\end{barticle}
\endbibitem

\bibitem[\protect\citeauthoryear{{Nazaryan} et~al.}{2012}]{N12}
\begin{barticle}
\bauthor{\bsnm{{Nazaryan}}, \binits{T.A.}},
\bauthor{\bsnm{{Petrosian}}, \binits{A.R.}},
\bauthor{\bsnm{{Mclean}}, \binits{B.J.}}:
\bjtitle{Astrophysics}
\bvolume{55},
\bfpage{448}
(\byear{2012})
\end{barticle}
\endbibitem

\bibitem[\protect\citeauthoryear{{Patton} et~al.}{2000}]{patton00}
\begin{barticle}
\bauthor{\bsnm{{Patton}}, \binits{D.R.}},
\bauthor{\bsnm{{Carlberg}}, \binits{R.G.}},
\bauthor{\bsnm{{Marzke}}, \binits{R.O.}},
\bauthor{\bparticle{et} \bsnm{al.}}:
\bjtitle{\apj}
\bvolume{536},
\bfpage{153}
(\byear{2000})
\end{barticle}
\endbibitem

\bibitem[\protect\citeauthoryear{{Patton} et~al.}{2011}]{patton11}
\begin{barticle}
\bauthor{\bsnm{{Patton}}, \binits{D.R.}},
\bauthor{\bsnm{{Ellison}}, \binits{S.L.}},
\bauthor{\bsnm{{Simard}}, \binits{L.}},
\bauthor{\bparticle{et} \bsnm{al.}}:
\bjtitle{\mnras}
\bvolume{412},
\bfpage{591}
(\byear{2011})
\end{barticle}
\endbibitem

\bibitem[\protect\citeauthoryear{{Patton} et~al.}{2013}]{patton13}
\begin{botherref}
\oauthor{\bsnm{{Patton}}, \binits{D.R.}},
\oauthor{\bsnm{{Torrey}}, \binits{P.}},
\oauthor{\bsnm{{Ellison}}, \binits{S.L.}},
\bauthor{\bparticle{et} \bsnm{al.}}:
ArXiv:
\arxivurl{1305.1595}
(2013)
\end{botherref}
\endbibitem

\bibitem[\protect\citeauthoryear{{Petrosian} and {Turatto}}{1990}]{petrosian90}
\begin{barticle}
\bauthor{\bsnm{{Petrosian}}, \binits{A.R.}},
\bauthor{\bsnm{{Turatto}}, \binits{M.}}:
\bjtitle{\aap}
\bvolume{239},
\bfpage{63}
(\byear{1990})
\end{barticle}
\endbibitem

\bibitem[\protect\citeauthoryear{{Petrosian} and {Turatto}}{1995}]{petrosian95}
\begin{barticle}
\bauthor{\bsnm{{Petrosian}}, \binits{A.R.}},
\bauthor{\bsnm{{Turatto}}, \binits{M.}}:
\bjtitle{\aap}
\bvolume{297},
\bfpage{49}
(\byear{1995})
\end{barticle}
\endbibitem

\bibitem[\protect\citeauthoryear{{Petrosian} et~al.}{2005}]{petrosian05}
\begin{barticle}
\bauthor{\bsnm{{Petrosian}}, \binits{A.}},
\bauthor{\bsnm{{Navasardyan}}, \binits{H.}},
\bauthor{\bsnm{{Cappellaro}}, \binits{E.}},
\bauthor{\bparticle{et} \bsnm{al.}}:
\bjtitle{\aj}
\bvolume{129},
\bfpage{1369}
(\byear{2005})
\end{barticle}
\endbibitem

\bibitem[\protect\citeauthoryear{{Petrosian} et~al.}{2007}]{P07}
\begin{barticle}
\bauthor{\bsnm{{Petrosian}}, \binits{A.}},
\bauthor{\bsnm{{McLean}}, \binits{B.}},
\bauthor{\bsnm{{Allen}}, \binits{R.J.}},
\bauthor{\bsnm{{MacKenty}}, \binits{J.W.}}:
\bjtitle{\apjs}
\bvolume{170},
\bfpage{33}
(\byear{2007})
\end{barticle}
\endbibitem

\bibitem[\protect\citeauthoryear{{Petrosian} et~al.}{2008}]{P08}
\begin{barticle}
\bauthor{\bsnm{{Petrosian}}, \binits{A.}},
\bauthor{\bsnm{{McLean}}, \binits{B.}},
\bauthor{\bsnm{{Allen}}, \binits{R.}},
\bauthor{\bparticle{et} \bsnm{al.}}:
\bjtitle{\apjs}
\bvolume{175},
\bfpage{86}
(\byear{2008})
\end{barticle}
\endbibitem

\bibitem[\protect\citeauthoryear{{Schlafly} and
  {Finkbeiner}}{2011}]{schlafly11}
\begin{barticle}
\bauthor{\bsnm{{Schlafly}}, \binits{E.F.}},
\bauthor{\bsnm{{Finkbeiner}}, \binits{D.P.}}:
\bjtitle{\apj}
\bvolume{737},
\bfpage{103}
(\byear{2011})
\end{barticle}
\endbibitem

\bibitem[\protect\citeauthoryear{{Schlegel} et~al.}{1998}]{schlegel98}
\begin{barticle}
\bauthor{\bsnm{{Schlegel}}, \binits{D.J.}},
\bauthor{\bsnm{{Finkbeiner}}, \binits{D.P.}},
\bauthor{\bsnm{{Davis}}, \binits{M.}}:
\bjtitle{\apj}
\bvolume{500},
\bfpage{525}
(\byear{1998})
\end{barticle}
\endbibitem

\bibitem[\protect\citeauthoryear{{Shaw}}{1979}]{shaw79}
\begin{barticle}
\bauthor{\bsnm{{Shaw}}, \binits{R.L.}}:
\bjtitle{\aap}
\bvolume{76},
\bfpage{188}
(\byear{1979})
\end{barticle}
\endbibitem

\bibitem[\protect\citeauthoryear{{Smartt}}{2009}]{smartt09}
\begin{barticle}
\bauthor{\bsnm{{Smartt}}, \binits{S.J.}}:
\bjtitle{\araa}
\bvolume{47},
\bfpage{63}
(\byear{2009})
\end{barticle}
\endbibitem

\bibitem[\protect\citeauthoryear{{Smith} et~al.}{2007}]{smith07}
\begin{barticle}
\bauthor{\bsnm{{Smith}}, \binits{B.J.}},
\bauthor{\bsnm{{Struck}}, \binits{C.}},
\bauthor{\bsnm{{Hancock}}, \binits{M.}},
\bauthor{\bparticle{et} \bsnm{al.}}:
\bjtitle{\aj}
\bvolume{133},
\bfpage{791}
(\byear{2007})
\end{barticle}
\endbibitem

\bibitem[\protect\citeauthoryear{{Tabachnick} and {Fidell}}{2006}]{tabachnick06}
\begin{bbook}
\bauthor{\bsnm{{Tabachnick}}, \binits{B.G.}},
\bauthor{\bsnm{{Fidell}}, \binits{L.S.}}:
\bbtitle{{Using multivariate statistics}},
\bedition{5}nd edn.
\bpublisher{Allyn \& Bacon},
\blocation{Needham Heights, MA, USA}
(\byear{2006})
\end{bbook}
\endbibitem

\bibitem[\protect\citeauthoryear{{Terry} et~al.}{2002}]{terry02}
\begin{barticle}
\bauthor{\bsnm{{Terry}}, \binits{J.N.}},
\bauthor{\bsnm{{Paturel}}, \binits{G.}},
\bauthor{\bsnm{{Ekholm}}, \binits{T.}}:
\bjtitle{\aap}
\bvolume{393},
\bfpage{57}
(\byear{2002})
\end{barticle}
\endbibitem

\bibitem[\protect\citeauthoryear{{van den Bergh}}{1997}]{vandenbergh97}
\begin{barticle}
\bauthor{\bsnm{{van den Bergh}}, \binits{S.}}:
\bjtitle{\aj}
\bvolume{113},
\bfpage{197}
(\byear{1997})
\end{barticle}
\endbibitem

\bibitem[\protect\citeauthoryear{{van den Bergh} et~al.}{2005}]{vandenbergh05}
\begin{barticle}
\bauthor{\bsnm{{van den Bergh}}, \binits{S.}},
\bauthor{\bsnm{{Li}}, \binits{W.}},
\bauthor{\bsnm{{Filippenko}}, \binits{A.V.}}:
\bjtitle{\pasp}
\bvolume{117},
\bfpage{773}
(\byear{2005})
\end{barticle}
\endbibitem

\bibitem[\protect\citeauthoryear{{Wang} et~al.}{2010}]{wang10}
\begin{barticle}
\bauthor{\bsnm{{Wang}}, \binits{J.}},
\bauthor{\bsnm{{Deng}}, \binits{J.S.}},
\bauthor{\bsnm{{Wei}}, \binits{J.Y.}}:
\bjtitle{\mnras}
\bvolume{405},
\bfpage{2529}
(\byear{2010})
\end{barticle}
\endbibitem

\end{thebibliography}

\end{document}